# TURBULENT MIXING, DIFFUSION AND GRAVITY IN THE FORMATION OF COSMOLOGICAL STRUCTURES: THE FLUID MECHANICS OF DARK MATTER

**Carl H. GIBSON**

Departments of AMES and SIO, University of California at San Diego
La Jolla, CA 92093-0411, USA
E-mail: *cgibson@ucsd.edu*

## ABSTRACT

The theory of gravitational structure formation in astrophysics and cosmology is revised based on real fluid behavior and turbulent mixing theory. Gibson's 1996-1998 theory balances fluid mechanical forces with gravitational forces and density diffusivity with gravitational diffusivity at critical viscous, turbulent, magnetic, and diffusion length scales termed Schwarz scales $L_{SX}$. Condensation and void formation occurs on non-acoustic density nuclei produced by turbulent mixing for scales $L \geq L_{SX_{max}}$ rather than on sound wave crests and troughs for $L \geq L_J$ as required by Jeans's 1902 linear acoustic theory. Schwarz scales $L_{SX} = L_{SV}, L_{ST}, L_{SM}$, or $L_{SD}$ may be smaller or larger than Jeans's scale $L_J$. Thus, a very different "nonlinear" cosmology emerges to replace the currently accepted "linear" cosmology. According to the new theory, most of the inner halo dark matter of galaxies consists of planetary mass objects that formed soon after the plasma to neutral gas transition 300,000 years after the Big Bang. These objects are termed primordial fog particles (PFPs) and provide an explanation for Schild's 1996 "rogue planets ... likely to be the missing mass" of his observed quasar-lens galaxy, inferred from the twinkling frequencies of both quasar images and their phased difference. The more massive nonbaryonic dark matter (possibly neutrinos) is super-diffusive because of its small collisional cross-section with ordinary (baryonic) matter, and can only condense at $L_{SD}$ scales much larger than galaxies to form massive halos of galaxy superclusters, clusters and outer galaxy halos. In the beginning of structure formation 30,000 years after the Big Bang, viscous Schwarz scales $L_{SV}$ matched the Hubble scale $ct$ of causal connection at protosupercluster masses of $10^{46}$ kg, with photon viscosity values of $5 \times 10^{26}$ m$^2$ s$^{-1}$, where c is the velocity of light and t is the age of the universe, decreasing to $10^{41}$ kg protogalaxy masses at plasma neutralization. Diffusivities of $10^{28}$ m$^2$ s$^{-1}$ are indicated by clusterhalo $L_{SD}$ values of $10^{22}$ m, with nonbaryonic dark matter dominating luminous matter by a factor of about 800, as observed in a dense galaxy cluster by Tyson and Fischer (1995).

## NOMENCLATURE

AU = astronomical unit, $1.4960 \times 10^{11}$ m
$a(t)$ = cosmological scale factor as a function of time t
c = speed of light, $2.9979 \times 10^{8}$ m s$^{-1}$
D = molecular diffusivity of density, m$^2$ s$^{-1}$
$\varepsilon$ = viscous dissipation rate, m$^2$ s$^{-3}$
G = Newton's gravitational constant, $6.7 \times 10^{-11}$ m$^3$ kg$^{-1}$ s$^{-2}$
$\gamma$ = rate of strain, s$^{-1}$
$k_B$ = Boltzmann's constant, $1.38 \times 10^{-23}$ J K$^{-1}$
l = collision length, m
ly = light year, $9.461 \times 10^{15}$ m
$L_{GIV}$ = gravitational-inertial-viscous scale, $[\varepsilon^2/\rho G^3]^{1/4}$
$L_{SV}$ = viscous Schwarz scale, $(\gamma \nu / \rho G)^{1/2}$
$L_{ST}$ = turbulent Schwarz scale, $\varepsilon^{1/2}/(\rho G)^{3/4}$
$L_{SD}$ = diffusive Schwarz scale, $(D^2/\rho G)^{1/4}$
$L_H$ = Hubble or horizon scale of causal connection, ct
$L_J$ = Jeans scale, $V_S/(\rho G)^{1/2}$

$\lambda$ = wavelength, m
$m_p$ = proton mass, $1.661 \times 10^{-27}$ kg
$M_{sun}$ = solar mass, $1.99 \times 10^{30}$ kg
$M_{earth}$ = earth mass, $5.977 \times 10^{24}$ kg
$\nu$ = kinematic viscosity, $m^2 s^{-1}$
pc = parsec, $3.0856 \times 10^{16}$ m
R = gas constant, $m^2 s^{-2} K^{-1}$
R(t) = cosmological scale, $a(t) = R(t)/R(t_0)$
$\rho$ = density, kg $m^{-3}$
$\rho_C$ = critical density, $10^{-26}$ kg $m^{-3}$ at present for flat universe
$\sigma$ = collision cross section, $m^2$
$\sigma_T$ = Thomson cross section, $6.6524 \times 10^{-29}$ $m^2$
t = time since Big Bang
$t_0$ = present time, $4.6 \times 10^{17}$ s
T = temperature, K
$V_S$ = sound speed, m $s^{-1}$
z = redshift = $\lambda/\lambda_0 - 1$

## INTRODUCTION

The Jeans (1902) theory of gravitational instability fails to correctly describe this highly nonlinear phenomenon because it is based on a linear perturbation stability analysis of an inadequate set of conservation equations that exclude turbulence, turbulent mixing, viscous forces, and molecular and gravitational diffusivity. Linear theories typically give vast errors when applied to nonlinear processes. For example, neglect of the inertial-vortex forces in the Navier Stokes equations gives laminar flow velocity profiles that are independent of Reynolds number, but these profiles are contrary to observations that such flows always become turbulent, with vastly different stresses and velocity fields, when the Reynolds number exceeds a critical value. It is argued by Gibson (1996) that the dark matter paradox is one of several cosmological misconceptions resulting from the application of Jeans's gravitational instability criterion to the development of structure by gravitational forces in the early universe.

Recently, information about the early universe has been flooding in from bigger and better telescopes on the earth's surface, and a host of space telescopes covering spectral bands previously unobservable, all with spectacular resolution. They reveal that structure existed much earlier and at both larger and smaller scales than expected from cosmological models such as Weinberg (1972), Zel'dovich and Novikov (1983), Silk (1989, 1994), Kolb and Turner (1993), Peebles (1993), and Padmanabhan (1993) that rely on Jeans's theory. Predictions from Jeans's linear theory will be termed "linear cosmology" in contrast to "nonlinear cosmology" based on the Gibson (1996) Schwarz scale fluid mechanical criteria.

Jeans's (1902) theory neglects viscous and nonlinear terms in the Navier Stokes momentum equations, reducing the problem of gravitational instability in a nearly uniform gas to one of linear acoustics. Sound waves of wavelength $\lambda$ require a time $\lambda/V_S$ to propagate a distance of one wavelength, where $V_S$ is the speed of sound, and gravitational condensation requires a free fall time of $(\rho G)^{-1/2}$, where $\rho$ is the density and G is Newton's constant of gravitation. Jeans's criterion follows by setting these two times equal to each other, giving the Jeans acoustic length scale $L_J \equiv V_S/(\rho G)^{1/2}$. Sound waves provide density nuclei for gravitational condensation, but only for wavelengths $\lambda \geq L_J$. Therefore, the Jeans criterion for the gravitational instability of a density perturbation on scale L is

$$L \geq L_J \equiv V_S/(\rho G)^{1/2}.$$

Waves with $\lambda \leq L_J$ propagate away before gravity can act.

However, most density nuclei in natural fluids are non-acoustic, moving with the fluid velocities rather than the sound speed. They result from turbulent scrambling of the temperature and chemical species fluctuations that determine the density field $\rho$. The reference pressure fluctuation $\Delta p$ for sound in air is $2 \times 10^{-5}$ atmospheres, corresponding to isentropic $\Delta \rho / \rho$ of $6 \times 10^{-11}$ and $\Delta T/T$ of $1.4 \times 10^{-10}$. Measurements of $\Delta T/T$ for the cosmic microwave background (CMB) show $\Delta \rho / \rho$ values of about $10^{-5}$. If this fluctuation level were to occur as audible acoustic fluctuations in air on earth, it would represent a deafening sound level of 104 dB, close to the 125 dB threshold of pain. Since at the CMB time 300,000 years after the Big Bang there were only weak sources of sound, it seems certain that most of the density nuclei existing then were non-acoustic.

According to the turbulent mixing theory of Gibson (1968), constant density surfaces move with the local fluid velocity except for their velocity with respect to the fluid due to molecular diffusion. The scale of the smallest density fluctuation is set by an equilibrium between the diffusion velocity D/L and the convection velocity $\gamma L$ at distances L away from points of maximum and minimum density, giving the Batchelor scale $L_B \equiv (D/\gamma)^{1/2}$ independent of the ratio $Pr \equiv \nu/D$, where $\gamma$ is the local rate-of-strain, D is the molecular diffusivity of $\rho$, and $\nu$ is the kinematic viscosity. This prediction has been confirmed by measurements in air, water and mercury for 0.02 ≤ Pr ≤ 700 and by numerical simulations at smaller Re for $10^{-2}$ ≤ Pr ≤ 1, Gibson et al. (1988). Even if gravitational condensation of mass were to take place on a sound wave moving in a stationary fluid, it would immediately produce a non-acoustic density maximum from the conservation of momentum. Since the ambient condensing fluid is not moving its momentum (zero) would immediately dominate the tiny momentum of the sound wave crest.

Gibson (1996) shows that gravitational condensation on a non-acoustic density maximum is limited by either viscous or turbulent forces at either the viscous Schwarz scale $L_{SV}$ ≡ $(\nu/\rho G)^{1/2}$ or the turbulent Schwarz scale $L_{ST}$ ≡ $\varepsilon^{1/2}/(\rho G)^{3/4}$, whichever is larger, where $\varepsilon$ is the viscous dissipation rate of the turbulence. For the superdiffusive non-baryonic dark matter that constitutes most of mass of the universe, the diffusive Schwarz scale $L_{SD}$ ≡ $[D^2/\rho G]^{1/4}$ limits condensation. The revised criterion for gravitational condensation at scale L is

$$L ≥ L_{SXmax} = \max [L_{SV}, L_{ST}, L_{SD}]$$

where only viscous and turbulent forces are assumed to prevent condensation in the early universe (magnetic forces are negligible) for the baryonic matter, and $L_{SD}$ sets the maximum scale for condensation of the non-baryonic matter ($\nu < [\varepsilon/\rho G]^{1/2}$).

$L_{SD}$ is derived by setting the diffusion velocity D/L equal to the gravitational velocity $L(\rho G)^{1/2}$. The diffusivity D of a gas is the particle collision length l times the particle velocity v. If l ≥ $L_H$ the particle is considered collisionless, and more complex methods are required using the collisionless Boltzmann equation and general relativity theory. Density perturbations in collisionless species like neutrinos are subject to Landau damping, also termed collisionless phase mixing or free streaming, Kolb and Turner (1993, p351). The free streaming length $L_{FS}$ is about $10^{24}$ m for neutrinos assuming a neutrino mass of $10^{-35}$ kg corresponding to that required for a flat universe, giving an effective diffusivity of $3 \times 10^{35}$ m$^2$ s$^{-1}$ from $L_{SD}$. Thus if neutrinos are the missing non-baryonic mass, they are irrelevant to structure formation until $L_{FS} = L_H$ at about $10^8$ years. Heavy "cold dark matter" non-baryonic particles become nonrelativistic early in the plasma epoch and condense at galactic scales, and give better fits to the observed structures from numerical N-body simulations of their "bottom up" gravitational clustering than do "hot dark matter" neutrinos.

In the early universe, the sound speed $V_S$ was very large because of the high temperatures, and horizon scale Reynolds numbers Re ≡ $c^2 t/\nu$ were small because the viscosity $\nu$ was large and t was small. Therefore, $L_{SV}$ and $L_{ST}$ were smaller than $L_J$, giving smaller mass condensations in this period of time. From linear cosmology, no condensation is possible in the plasma epoch following the Big Bang, with t ≤ 300,000 years ($10^{13}$ s) because $L_J > L_H$ ≡ ct, where c is the speed of light, t is the time, and $L_H$ is the scale of causal connection. No structures can form by causal processes on scales larger than $L_H$ because the speed of information transfer is limited by the speed of light. Star formation is prevented by the Jeans criterion until the Jeans mass $M_J = (RT/\rho G)^{3/2}$ decreases below a solar mass as the temperature of the universe decreases, but this requires hundreds of millions of years rather than only a few million years by the present theory. Recent observations suggest that stars, galaxies and galaxy clusters existed at the earliest times observable; that is, at times t much less than a billion years with redshifts z of 4 and larger.

No viscous or turbulent limitations prevent condensation after about 30,000 years (t = $10^{12}$ s) when decreasing $L_{SV}$ values first matched the increasing horizon scale $L_H$ with rate of strain $\gamma = 1/t$ and $\nu$ values more than $10^{26}$ m$^2$ s$^{-1}$, Gibson (1997ab). At this time the horizon mass $L_H^3 \rho$ equaled the Schwarz viscous mass $M_{SV} = L_{SV}^3 \rho$ at the observed supercluster mass of $10^{46}$ kg, Kolb and Turner (1993), the largest structure in the universe. Density as a function of time can be computed from Einstein's equations of general relativity assuming a flat universe (kinetic energy matching gravitational potential energy), Weinberg (1972, Table 15.4). The horizon Reynolds number $c^2 t/\nu$ therefore was about 150, near transition.

This enormous kinematic viscosity can be explained as due to photon collisions with electrons of the plasma of H and He ions by Compton scattering, with Thomson cross section $\sigma_T = 6.7 \times 10^{-29}$ m$^2$. Fluctuations of plasma velocity are smoothed by the intense radiation since the ions remain closely coupled to the electrons by electric forces. We estimate the kinematic viscosity $\nu$ to be lc = $5 \times 10^{26}$ m$^2$ s$^{-1}$ using a collision length l of $10^{18}$ m from l = 1/$\sigma_T$n, with number density n of electrons about $10^{10}$ m$^{-3}$ assuming the baryon (ordinary) matter density is $10^{-2}$ less than the critical density $\rho_C = 10^{-15}$ kg m$^{-3}$ at the time ($\rho_C = 10^{-26}$ at present), from Weinberg (1972).

Between 30,000 years and 300,000 years during the plasma epoch of the universe the temperature decreased from $10^5$ to 3000 K, decreasing the viscosity for the baryonic matter if its expansion were gravitationally arrested ( const), and decreasing the viscous Schwarz scale $L_{SV}$ of condensation due to decreases in both and . The final condensation mass by this scenario is about $10^{41}$ kg, the mass of a large galaxy. If the expansion were merely decelerated rather than arrested, the proto-galaxy mass would increase because the density decreases. True condensation with increasing plasma density would produce proto-galaxies of smaller mass. In all cases the Reynolds number should be marginally subcritical, which is consistent with the cosmic microwave background observations of extremely uniform temperature of 2.735 K, with fluctuation T/T about $10^{-5}$. If the flow were strongly turbulent in the primordial plasma, T/T values would be larger by about 3-4 orders of magnitude, thus supporting the conclusion that the primordial plasma was not strongly turbulent before the transition to neutral gas because viscous forces were larger than the inertial vortex forces. Any gravitational structure formation during the plasma epoch would also result in turbulence suppression, by buoyancy forces within the structures as shown by Gibson (1988).

Because the non-baryonic matter is decoupled from the baryonic plasma by lack of any collisional mechanisms, it should fill the expanding voids between the proto-superclusters, proto-clusters, and proto-galaxies that developed during the plasma epoch, Gibson (1996). The average density of galaxies today is ten times less than the protogalactic baryonic density of $10^{-20}$ kg m$^{-3}$, so it appears that the effect of gravity was to decelerate rather than arrest their expansion. A baryonic density of $10^{-17}$ kg m$^{-3}$ matches the density of globular star clusters, which may be no coincidence. This was the baryonic density at about 10,000 years when mass first matched energy, which may have been preserved as a remnant. At some point gravitational forces caused condensation of the non-baryonic matter as well, as halos of the evolving baryonic structures with galaxy to supercluster masses.

From the $6 \times 10^{21}$ m thickness scale of the dark matter halo of a dense galaxy cluster, computed for the first time by tomography from gravitational arcs of thousands of background galaxies by Tyson and Fischer (1995), a diffusivity D of order $10^{28}$ m$^2$ s$^{-1}$ may be inferred by setting the scale equal to the Schwarz diffusive scale $L_{SD}$. This is too large by $10^{13}$ to be baryonic matter, so we can conclude that the cluster halo consists of non-baryonic matter with this large effective diffusivity. It follows that the dark matter of at least the inner core of galaxies is likely to be mostly baryonic, since the non-baryonic component diffuses away to form halos on the larger structures.

Because the Jeans criterion will not permit baryonic matter to condense to form the observed structures, standard linear cosmology requires the non-baryonic dark matter to condense early in the plasma epoch, forming gravitational potential wells to guide the condensation of the baryonic matter. This is accomplished by assuming the weakly interacting massive particles (WIMPs) have large masses, about $10^{-25}$ kg, giving, perforce, small particle velocities and small initial condensation masses in the galaxy mass range. Ad hoc mixtures of such "cold dark matter" with less massive "warm" and "hot" dark matter particles, plus adjustments to the cosmological constant are required to match observations of the actual universe structure. Moreover, it is necessary to assume an open universe with about 20% of the critical density, and to resurrect the infamous cosmological constant , first introduced by Einstein and later renounced. Such unconvincing curve fitting is no longer required if the Jeans criterion is abandoned in favor of the recommended Schwarz length scale fluid mechanical criteria. If the universe is open and a nonzero value of is required to make it flat, the effect is to reduce the nonbaryonic component.

In the following we first review the theory of gravitational condensation. The linear acoustic Jeans theory is discussed, and replaced by a nonlinear theory based on the mechanics of real fluids. Cosmological differences between the theories are reviewed, and comparisons are made with observations. Finally, a summary and conclusions are provided.

## THEORY OF GRAVITATIONAL CONDENSATION

Gravitational condensation for scales smaller than the horizon $L_H$ in the early universe can be described by the Navier Stokes equations of momentum conservation

$$\frac{\partial \vec{v}}{\partial t} = -\nabla B + \vec{v} \times \vec{\omega} + \vec{F}_g + \vec{F}_v + \vec{F}_m + \vec{F}_{etc.} \qquad (1)$$

where $\vec{v}$ is the velocity, $B = p/\rho + v^2/2$ is the Bernoulli group, $\vec{v} \times \vec{\omega}$ is the inertial vortex force, $\vec{\omega}$ is the vorticity, $\vec{F}_g$ is the gravitational force, $\vec{F}_v$ is the viscous force, and the magnetic and other forces $\vec{F}_m + \vec{F}_{etc.}$ are assumed to be negligible. The gravitational force per unit mass $\vec{F}_g = -\nabla\phi$, where $\phi$ is the gravitational potential in the expression

$$\nabla^2 \phi = 4\pi G \rho \qquad (2)$$

in a fluid of density $\rho$. The density conservation equation in the vicinity of a density maximum or minimum is

$$\frac{\partial \rho}{\partial t} + \vec{v} \cdot \nabla\rho = D_{eff} \nabla^2 \rho \qquad (3)$$

where the effective diffusivity $D_{eff}$

$$D_{eff} = D - L^2/\tau_g \qquad (4)$$

includes the molecular diffusivity D of the gas and a negative gravitational term depending on the distance $L \leq L_{SX_{max}}$ from the non-acoustic density nucleus and the gravitational free fall time $\tau_g = (\rho G)^{-1/2}$. Turbulence is driven by $\vec{v} \times \vec{\omega}$ forces.

For scales smaller than $L_H$ gravitational effects are described by Einstein's nonlinear equations of general relativity

$$R_{ij} - g_{ij}R = -8\pi G T_{ij} \qquad (5)$$

where $R_{ij}$ is the Ricci tensor, R is its trace, a term $\Lambda g_{ij}$ on the right has been set to zero (the cosmological constant $\Lambda$ was arbitrarily introduced by Einstein to prevent expansion of the universe and later dropped), G is Newton's gravitational constant, $T_{ij}$ is the energy-momentum tensor, $g_{ij}$ is the metric tensor, and indices i and j are 0, 1, 2, and 3. The nonlinearity represents the effects of gravity on itself. The Ricci tensor is formed by contracting the fourth order curvature tensor so that the gravitational field tensor on the left of (5) has only terms linear in second derivatives or quadratic in first derivatives of the metric tensor, Weinberg (1972, p153). It was developed to account for curvature problems of non-Euclidean geometry by Riemann and Christoffel and was adapted by Einstein to preserve Lorentz invariance and the equivalence of inertia and gravitation in mechanics and electromechanics, Weinberg (1972). Full discussion of classical solutions of the Einstein gravitational field equations are given by standard cosmology texts such as Weinberg (1972), Peebles (1993), Kolb and Turner (1993), and Padmanabhan (1993).

The homogeneous, isotropic Robertson-Walker metric is assumed to describe the universe after the Big Bang, where the cosmic scale factor $a(t) = R(t)/R(t_0)$ measures the time evolution of spatial scales in comoving coordinates as the universe expands to the present time $t_0$. Variations in curvature of space can result in acausal increases of density for scales larger than $L_H$. Isocurvature fluctuations may not grow after inflationary expansion beyond the horizon, and reenter the horizon at a later time with the same amplitude, Kolb and Turner (1993). Curvature fluctuations grow with the cosmological scale $R(t) \sim t^{2/3}$ until they reenter the horizon.

## LINEAR THEORY

Jeans (1902) considered the problem of gravitational condensation in a stagnant, inviscid gas with small perturbations of density, potential, pressure, and velocity so that the nonlinear term in (1) could be neglected along with all other terms except $\vec{F}_g$. He assumed that the pressure p is a function only of the density $\rho$. Either the linear perturbation assumptions of Jeans or his barotropic assumption are sufficient to reduce the problem to one of acoustics. Details of the Jeans derivation are given in Kolb and Turner (1993, 342-344) and in most other standard textbooks on cosmology, so they will not be repeated here. Diffusion terms are neglected in equation (3), and the adiabatic sound speed $V_S = \sqrt{\partial p/\partial \rho}$ is taken from the assumption that there are no variations in the equation of state. Cross differentiation with respect to space and time of the perturbed equations neglecting second order terms gives a wave equation for the density perturbation $\rho_1$

$$\frac{\partial^2 \rho_1}{\partial t^2} - V_S^2 \nabla^2 \rho_1 = 4\pi G \rho_0 \rho_1 \qquad (6)$$

where $\rho_0$ is the unperturbed density. The solutions of (6) are of the form

$$\rho_1(\vec{r},t) = \rho(\vec{r},t) - \rho_0 = A \exp[-i\vec{k}\cdot\vec{r} + i\omega t] \rho_0 \qquad (7)$$

which are sound waves of amplitude A for large $k \gg k_J$ which obey a dispersion relation

$$\omega^2 = V_S^2 k^2 - 4\pi G \rho_0 \qquad (8)$$

where $k = |\vec{k}|$ and the critical wavenumber

$$k_J = (4\pi G \rho_0/V_S^2)^{1/2} \qquad (9)$$

has been interpreted as the criterion for gravitational instability. All solutions of (6) with wavelength larger than $L_J$ are imaginary and are termed gravitationally unstable in linear cosmologies. Only such modes are considered to be eligible for condensation to form structure. Void formation is very badly modeled by linear cosmologies, and is not even mentioned in standard treatments such as Kolb and Turner (1993).

## NONLINEAR THEORY

Consider the problem of gravitational instability for a nonacoustic density nucleus of diameter L and mass $M' = \rho L^3$, where $L_J > L > L_{SXmax}$. For scales smaller than $L_J$ the pressure adjusts rapidly compared to the gravitational time $\tau_g \approx (\rho G)^{-1/2}$. For scales larger than the largest Schwarz scale $L_{SXmax}$ fluid mechanical forces and molecular diffusion are negligible compared to gravitational forces toward or away from the nucleus. Starting from rest, we see that the system is absolutely unstable to gravitational condensation or void formation, depending on whether $M'$ is positive or negative.

The radial velocity $v_r$ will be negative or positive depending on the sign of $M'$, and will increase linearly with time since the gravitational acceleration at radius r from the center of the nucleus is constant, with value $-M'G/r^2$. Thus

$$v_r = -M'Gt/r^2 \tag{10}$$

shows the mass flux

$$dM'/dt = -\rho v_r 4\pi r^2 = M'\rho G t 4\pi \tag{11}$$

into or away from the nucleus is constant with radius. Integrating (11) gives

$$M'(t) = M'(t_0)\exp[2\pi\rho G t^2] = M'(t_0)\exp[2\pi(t/\tau_g)^2] \tag{12}$$

where $M'(t_0)$ is the initial mass of the density nucleus. The only place where the density changes appreciably is at the core of the nucleus. We can define the core radius $r_c$ as

$$r_c = -v_r''t = M'Gt^2/r_c^2, \tag{13}$$

where $r_c$ is the distance from which core material has fallen in time t, either to or from the core. The core mass change $M''$ is then

$$M'' = \rho r_c^3 = M'\rho G t^2 = M'(t_0)(t/\tau_g)^2\exp[2\pi(t/\tau_g)^2] \tag{14}$$

from (13) and (12).

Note that the velocity near the core becomes large for small r according to (10). This will produce turbulence at condensation nuclei with positive $M'$ for times t of order $\tau_g$. For void nuclei, the velocity of the rarefaction wave is limited by the sound speed.

The viscous Schwarz scale $L_{SV}$ is derived by setting the viscous force $F_V = \mu\gamma L^2$ at scale L equal to the gravitational force $F_g = G\rho L^3 \rho L^3/L^2$, so

$$L_{SV} \approx (\gamma/\rho G)^{1/2} \tag{15}$$

where $\gamma$ is the rate of strain. Viscous forces overcome gravitational forces for scales smaller than $L_{SV}$. The turbulent Schwarz scale $L_{ST}$ is derived by setting the inertial vortex forces of turbulence $F_I = \rho V^2 L^2$ equal to $F_g = G\rho^2 L^4$, substituting the Kolmogorov expression $V = (\varepsilon L)^{1/3}$ for the velocity at scale L,

$$L_{ST} \approx \varepsilon^{1/2}/(\rho G)^{3/4} \tag{16}$$

where $\varepsilon$ is the viscous dissipation rate of the turbulence. These two scales become equal when the inertial, viscous, and gravitational forces coincide. The gravitational inertial viscous scale

$$L_{GIV} = [\nu^2/\rho G]^{1/4} \tag{17}$$

corresponds to this equality, where $L_{GIV} = L_{SD}$ if $D = \nu$.

We can compare these expressions with the Jeans scale

$$L_J \approx V_S/(\rho G)^{1/2} = [RT/\rho G]^{1/2} = [(p/\rho)/\rho G]^{1/2} \tag{18}$$

in terms of the temperature and pressure. The two forms for the sound velocity $V_S$ in (18) have led to the erroneous concepts of pressure support and thermal support, since by the Jeans criterion high temperature or pressure in a gas would prevent the formation of structure. The length scale $L_{IC} \approx [RT/\rho G]^{1/2}$ has the physical significance of an initial condensation scale in a uniform gas, based on the ideal gas law $p = \rho RT$, where increases in density are matched by increases in pressure so that the temperature remains constant. The length scale $L_{HS} \approx [(p/\rho)/\rho G]^{1/2}$ is a hydrostatic scale that arises if an isolated blob of gas approaches hydrostatic equilibrium, with zero pressure outside. Neither $L_{IC}$ nor $L_{HS}$ have any physical connection to the linear acoustic theory of Jeans (1902). $L_{IC}$ is reflected as the mass of globular clusters of stars: the initial fragmentation scale of protogalaxies emerging from the plasma epoch. $L_{HS}$ appears at the final stages of primordial fog particle formation as their size, but as an effect of the formation, not the cause.

## COSMOLOGY

The conditions of the primordial gas emerging from the plasma epoch are well specified. The composition was 75% hydrogen-1 and 25% helium-4 by mass, at a temperature of 3000 K. This gives a dynamical viscosity μ of $2.4\times10^{-5}$ kg m$^{-1}$ s$^{-1}$ by extrapolation of the mass averaged μ values for the components to 3000 K, so the kinematic viscosity $\nu = \mu/\rho$ depends on the density assumed. The most likely for the baryonic gas is $10^{-18}$ kg m$^{-3}$, which is the density existing at the time $10^{12}$ s of first structure formation. The plasma neutralization time was $10^{13}$ s, so the rate of strain $10^{-13}$ s$^{-1}$. This gives $L_{SV} = ( /G)^{1/2} = 2 \times 10^{14}$ m and $M_{SV} = L_{SV}^3 = 6.8 \times 10^{24}$ kg as the most likely mass of the first gravitationally condensing objects of the universe, termed primordial fog particles or PFPs. With this density the gravitational time $_g = (G)^{-1/2}$ is $10^{14}$ s, or 4 million years. However, the time required for the voids to isolate the individual PFPs should be much less since the speed of void boundaries represent rarefaction waves, and may thus approach the sound speed $V_S = 3 \times 10^3$ m s$^{-1}$, giving an isolation time $L_{SV}/V_S = 6 \times 10^{10}$ s, or only 2000 years. The viscous dissipation rate $= ^2 = 2 \times 10^{-13}$ m$^2$ s$^{-3}$. The range of estimated PFP masses for various densities and turbulence levels is $10^{23}$ to $10^{26}$ kg, in the small planetary range. Kolmogorov and Batchelor scales $L_K = L_B = 1.5 \times 10^{13}$ m.

Thus the entire baryonic universe of hydrogen and helium gas rapidly turned to fog as the cooling plasma universe neutralized, with resulting primordial fog particle masses approximately equal to that of the earth, separated by distances of over a thousand astronomical units. These PFPs constitute the basic materials of construction for everything else. Those that have failed to accrete to star mass, and this should be about 97%, constitute the baryonic dark matter. The mass of the inner halos of galaxies should be dominated by the mass of such PFPs, since the non-baryonic component diffuses to $L_{SD}$ scales that are much larger.

## OBSERVATIONS

### QUASAR MICROLENSING

Quasars are the most luminous objects in the sky. They are generally thought to represent black holes in cores of cannibal galaxies at an early stage of their formation when they were ingesting smaller galaxies, one or two billion years after the Big Bang. Quasar microlensing occurs when another galaxy is precisely on our line of sight to the quasar, so that it acts as a gravitational lens. The quasar image is split into two or more mirage-like images which twinkle at frequencies determined by the mass of the objects making up the lens galaxy. Schild (1996) reports the results of a 15 year study of the brightness fluctuations of the two images of the QSO Q0957+561 A,B gravitational lens, amounting to over 1000 nights of observations. The time delay of 1.1 years was determined to remove any effects of intrinsic quasar variability and the microlensing masses were determined by frequency analysis to be $10^{-5.5}$ solar masses, or $6.3 \times 10^{24}$ kg, precisely the same as the most likely primordial fog particle mass estimated above, and by Gibson (1996). Three observatories have since independently reported the same time delay and microlensing signals for this object. Thus it is an observational fact that the mass of at least one galaxy is dominated by planetary mass objects.

### PLANETARY NEBULA

Planetary nebula appear when ordinary stars are in a hot dying stage on their way to becoming white dwarfs. Strong stellar winds and intense radiation from the central star might cause ambient PFPs to reevaporate and reveal themselves. Hubble Space Telescope observations of the nearest planetary nebula Helix (NGC 7293), by O'Dell and Handron, reveal over 3500 "cometary knots" with mass about $3 \times 10^{25}$ kg that are strong PFP candidates, possibly the hypothetical objects which the authors describe as "comets brought out of cold storage". Thousands of similar "particles" also appear in HST photographs (PRC97-29, Sept. 18, 1997) of the recurring Nova T Pyxidis by M. Shara, R. Williams, and R. Gilmozzi.

### DENSE GALAXY CLUSTERS

Tyson and Fischer (1995) report the first mass profile of a dense galaxy cluster Abel 1689 from tomographic inversion of 6000 gravitational arcs of 4000 background galaxies. The mass of the cluster was $10^{45}$ kg, with density $5 \times 10^{-21}$ kg m$^{-3}$. From the reported mass contours the cluster halo thickness was about $6 \times 10^{21}$ m. Setting this size equal to $L_{SD} = [D^2/G]^{1/4}$ gives a diffusivity $D = 2 \times 10^{28}$ m$^2$ s$^{-1}$, which is much too large to be due to baryonic gas (by a factor of at least $10^{13}$).

## CONCLUSIONS

The Jeans acoustic criterion for gravitational instability should be completely abandoned. It is based on several faulty assumptions, and by the present theory overestimates the expected minimum mass of baryonic condensation by 2-5 orders of magnitude during the plasma epoch of the early universe when proto-supercluster to proto-galaxy objects were formed, and overestimates by 12 orders of magnitude during the following primordial gas epoch when primordial fog particles were formed. In the cold, dense, turbulent molecular clouds of present day galactic disks where most modern stars are being formed, the Jeans mass is likely to be smaller than the turbulent Schwarz mass by about an order of magnitude.

A fluid mechanically motivated criterion for gravitational condensation and void formation is recommended; that is

$$L \geq L_{SXmax} = \max [L_{SV}, L_{ST}, L_{SD}]$$

where structure formation occurs at scales L larger than the largest Schwarz scale.

According to the recommended criterion, gravitational structure formation in the universe began in the plasma epoch at a time about 30,000 years after the Big Bang with the formation of proto-supercluster-voids in the baryonic component. The fragmentation mass decreased to that of a proto-galaxy by the time of plasma neutralization at 300,000 years. Immediately following neutralization the baryonic fragmentation mass decreased to that of a small planet, so the universe of primordial hydrogen and helium gas turned to fog. The primordial fog particles have aggregated to form stars and everything else, but most are sequestered in galaxies as the dominant form of dark matter. The non-baryonic component of the universe, actually most of the matter, is superdiffusive because of its distinctive small cross section for collisions, and has diffused to $L_{SD}$ scales about $6 \times 10^{21}$ m that are larger by an order of magnitude than the inner halos of galaxies.